\documentclass[aps,prl,twocolumn,superscriptaddress]{revtex4-1} 
\usepackage{graphicx}
\usepackage{amsmath}

\begin{document}

\title{Single-mode subwavelength waveguides with wire metamaterials}

\author{Pavel A. Belov$^1$, Rost Dubrovka$^2$, Ivan Iorsh$^1$, Ilya Yagupov$^1$, and Yuri S. Kivshar$^{3}$}

\affiliation{National Research University for Information Technology, Mechanics and Optics (ITMO),
St.~Petersburg 197101, Russia\\
$^2$Queen Mary, University of London, London E1 4NS, UK\\
$^3$Nonlinear Physics Center, Research School of Physics and Engineering,
Australian National University, Canberra ACT 0200, Australia}

\begin{abstract}
We study the eigenmodes of a slab of a wire metamaterial and demonstrate
that such a waveguiding structure supports deep-subwavelength propagating modes
exhibiting properties of a single-mode waveguide at any fixed frequency below
the plasma frequency of metal wires. We compare our analytical results with
the dispersion relations extracted from the experimental measurements.
\end{abstract}

\maketitle

Guided electromagnetic waves are a special type of localized modes which exist either at interfaces of dissimilar media
(surface modes) or are trapped by the regions of high-dielectric material (waveguide modes). Guided modes can propagate in various
types of waveguides such as dielectric slab waveguides, photonic fibers, and metallic slot waveguides~\cite{WaveguideBook}.
Here we study a novel type of waveguide created by a slab of wire metamaterial~\cite{wirereview} of a finite thickness,
and exhibiting a number of peculiar features not found in conventional waveguides.

Wire metamaterials have a number of unique properties~\cite{wirereview}, including the possibility of the subwavelength image transfer~\cite{Subwavelength1,Subwavelength2,Subwavelength3}. Localized  modes in such structures have been studied
theoretically~\cite{Belovmodes,Theor}, and it was shown that these modes are similar to the so-called spoof plasmons, a special
class of surface modes which  propagate along corrugated metal or semiconductor surfaces~\cite{spoof001,spoof01,spoof}.
It was also shown that these guided modes can be useful for far-field superlensing \cite{Finkfirst,FinkPRL} and deep-subwavelength
waveguiding~\cite{Finkwaveguides} of electromagnetic waves.  However, despite of their peculiar properties, the studies of guided modes
and their possible applications were mainly theoretical with no experimental verification of their actual propagation in wire
metamaterials.

In this Letter,  we discuss several unique properties of this novel type of guided modes, compare them with the corresponding
modes of conventional dielectric waveguides, and also present the first direct experimental measurements of the guided modes
propagation in a slab of wire metamaterial.

\begin{figure}[!h]
\centerline{\includegraphics[width=0.8\columnwidth]{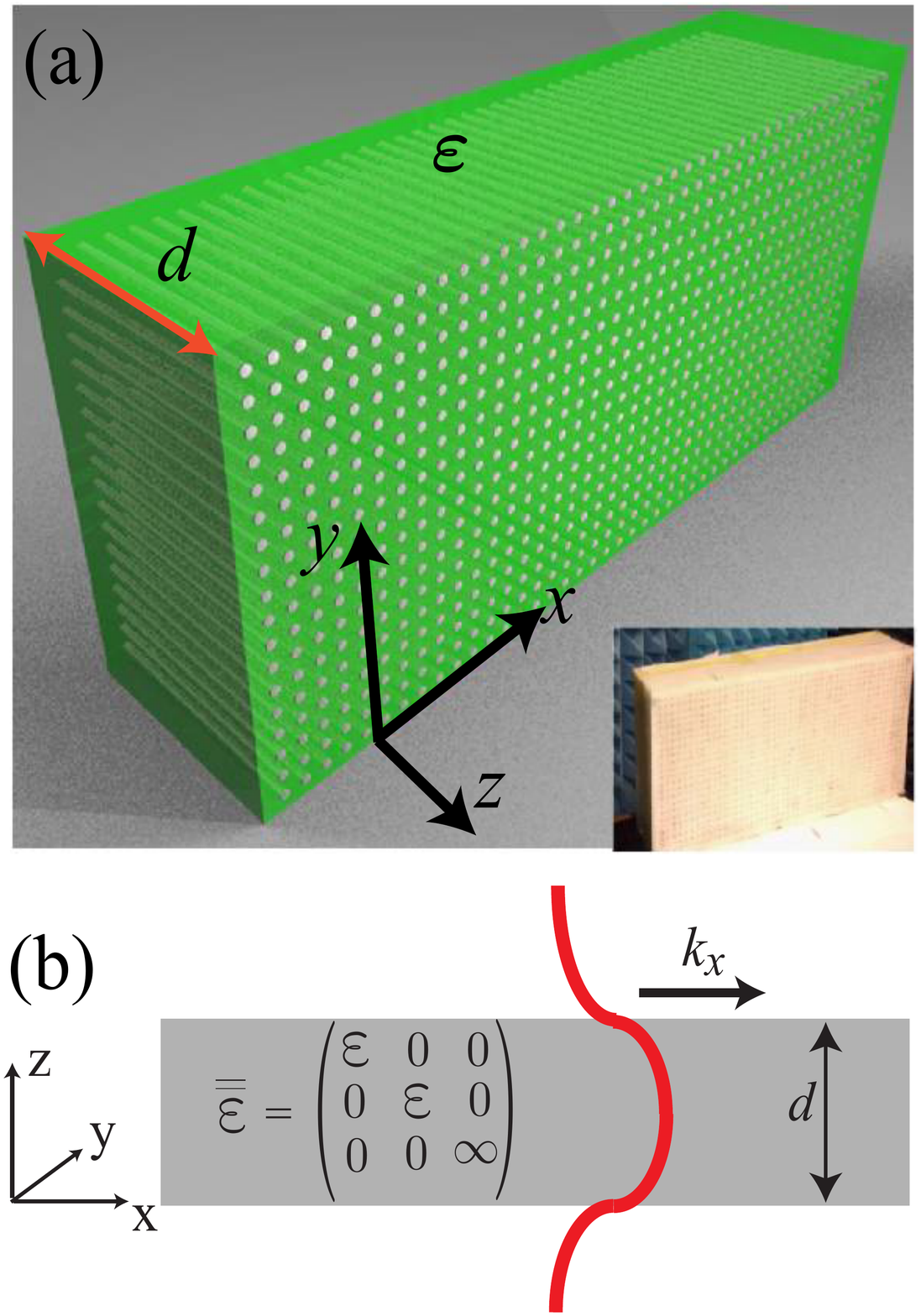}}
\caption{{(a) Geometry of a wire-metamaterial waveguide. Inset shows a photograph
of the experimental sample. (b) Propagation of the waveguide mode in the wire-metamaterial slab.}}
\label{fig0}
\end{figure}

We study the waveguiding wire-metamaterial structure shown in Fig.~\ref{fig0}(a) where guided modes propagate in the $x$ direction.
We describe a slab of wire metamaterial within a local effective theory approach. This approximation is valid when the wire-lattice period
is much less than the wavelength. Within this approximation, the wire metamaterial can be characterized by a diagonal tensor of dielectric permittivity
 $\bar{\bar{\varepsilon}}$ with one component parallel to the wires being equal to infinity, and two other components being equal to the dielectric
 permittivity $\varepsilon=n_m^2$ of the host medium~\cite{wirereview}. Next, we derive the dispersion relations for the waveguide modes of a slab waveguide
 made of this extremely anisotropic medium (see Fig.~\ref{fig0}(b)).

 To obtain analytical expressions for waveguide modes, we consider TM-polarization of light. We then recall that in the case of extremely anisotropic metamaterial it support propagation only of TEM-polarized waves with magnetic field given by $H_y= Ae^{in_m k_0 z}+Be^{-in_m k_0 z}$, where $k_0$ is the wavevector in vacuum. We then apply the continuity boundary conditions for the tangential magnetic and electric fields. Equating the determinant of the obtained linear system to zero, we find the dispersion equation for the localized modes~(see also Ref.~\cite{FinkWCM}),
\begin{align}
k_x = \left\{
     \begin{array}{lr}
       (k_0/n_m) [\tan^2(n_m k_0 d/2) + n_m^2]^{1/2},  & \eta > 0, \\
       (k_0/n_m) [\cot^2(n_m k_0 d/2) + n_m^2]^{1/2}  &  \eta < 0,
     \end{array}
   \right.  \label{disprel}
\end{align}
where $\eta \equiv \tan(n_m k_0 d/2)$. When the dielectric host medium is absent ($n_m=1$), Eqs.~\eqref{disprel}  become equivalent to the dispersion
relations obtained earlier in Ref.~\cite{Finkfirst}.

\begin{figure}[!h]
\centerline{\includegraphics[width=0.9\columnwidth]{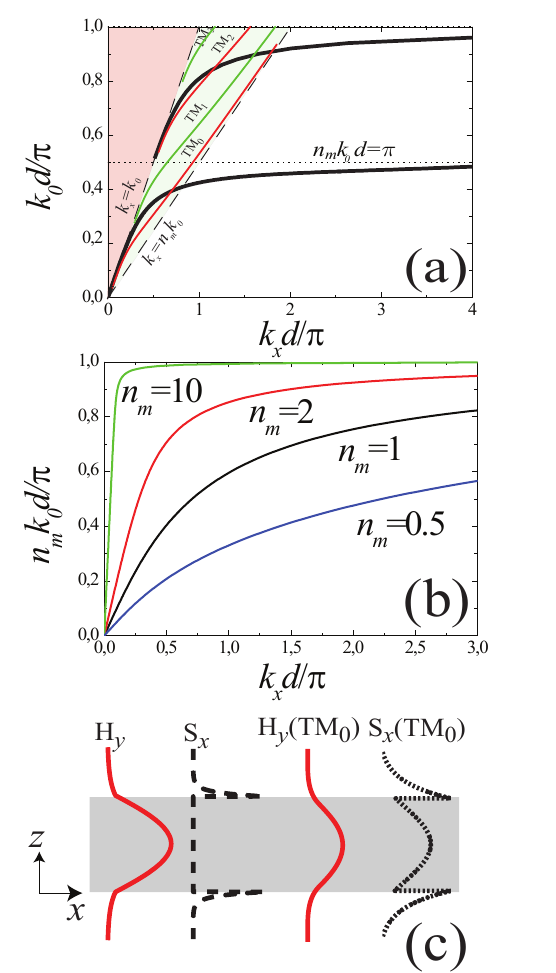}}
\caption{(a) Dispersion of the guided modes of a wire media slab (thick solid curves). 
Refractive index of the host media is 2.0. Pink area shows the region of the propagating 
modes in free space, blue area shows the region of the existence of the waveguide modes of the corresponding
conventional dielectric waveguide with the refractive index 2.0. Thin red and green lines show the dispersion 
of the waveguide modes in the dielectric waveguide. (b) Dispersion of the localized modes of a metamaterial 
slab for different values of refractive index of the host medium. (c) Profile of the magnetic field of the first  
mode of the wire media slab. Black dashed line shows the profile of the $x$ component of the Poynting vector, 
and a blue dotted line shows the Poynting vector of the corresponding dielectric waveguide mode. } 
\label{figdisp1}
\end{figure}

Figure~\ref{figdisp1}(a) presents the dispersion curves of the guided modes in a slab of wire metamaterial. For comparison, we show also the dispersion curves
of the guided modes in a slab of conventional dielectric material of the same thickness with the refractive index 2.0 [see red and green curves in Fig.~\ref{figdisp1}(a)].
We observe some similarities between the dispersions of dielectric and metamaterial waveguides. In particularly, both the structures have only one fundamental 
eigenmode with no cut-off. Higher-order modes are defined by the cut-off frequencies: $n_mk_0d=\pi n$, for the metamaterial waveguide, and $\sqrt{n_m^2-1}k_0d\approx \pi n$,
 for the dielectric waveguide, where $n$ is the index of the corresponding eigenmode. At the same time, we observe sufficient differences between the modes of 
 dielectric and metamaterial waveguides. First, the waveguide number of the dielectric waveguide is always limited with the refractive index of
 the waveguide core: $k_x<n_m k_0$, thus dispersion curves of all the guided modes lie in the blue area shown in Fig.~\ref{figdisp1}(a). 
 The waveguide number of the eigenmodes in the metamaterial waveguide generally is not bounded if we do not account for losses and internal periodicity of
the metamaterial, and they can extend to infinity, as in the case of surface plasmon polaritons. Moreover, only one eigenmode can exist in the metamaterial waveguide 
at the fixed frequency, whereas in the case of dielectric waveguide,  arbitrary large number of modes may exist at the fixed frequency for sufficiently large frequencies. 
Therefore, the metamaterial slab is a single-mode waveguide for arbitrary frequency (providing that the frequency is below the plasma frequency of the metal).
 While the studied eigenmodes exhibit both the properties of guided modes and surface waves, we call them the waveguide modes, because their dispersion depends
 on the thickness of the metamaterial slab, and the electric field is localized in the bulk rather than at the surface.

Figure~\ref{figdisp1}(b) shows the dispersion curves of the fundamental guided mode of the metamaterial slab depending on the refractive index of the host media. 
We notice that the guided mode exists even for the case when the refractive index is less than unity, which is not the case for dielectric waveguides, where 
no guided modes are possible for a low-index core. We should also mention that for the metamaterial eigenmodes the energy is transferred only outside 
the slab, and no energy transfer is present inside the slab, as shown in Fig.~\ref{figdisp1}(c).

Next, we study the waveguide modes experimentally for the microwave frequency range, using a sample made of a wired metamaterial shown in Fig.~\ref{fig0}. 
A two-dimensional array of metallic wires of length $10$~cm and period of $1$~cm is placed in a dielectric host media. The slab dimensions are $50\times25\times10$ cm. 
We excite the waveguiding metamaterial structure with a point source, positioned at the distance of $3$~mm from the structure surface. The signal is then collected 
with a probe at the opposite side of the structure.  Scanning the backside of the structure with the receiver allows to build the amplitude and phase maps 
of the transmitted radiation.

\begin{figure}[h]
\centerline{\includegraphics[width=0.95\columnwidth]{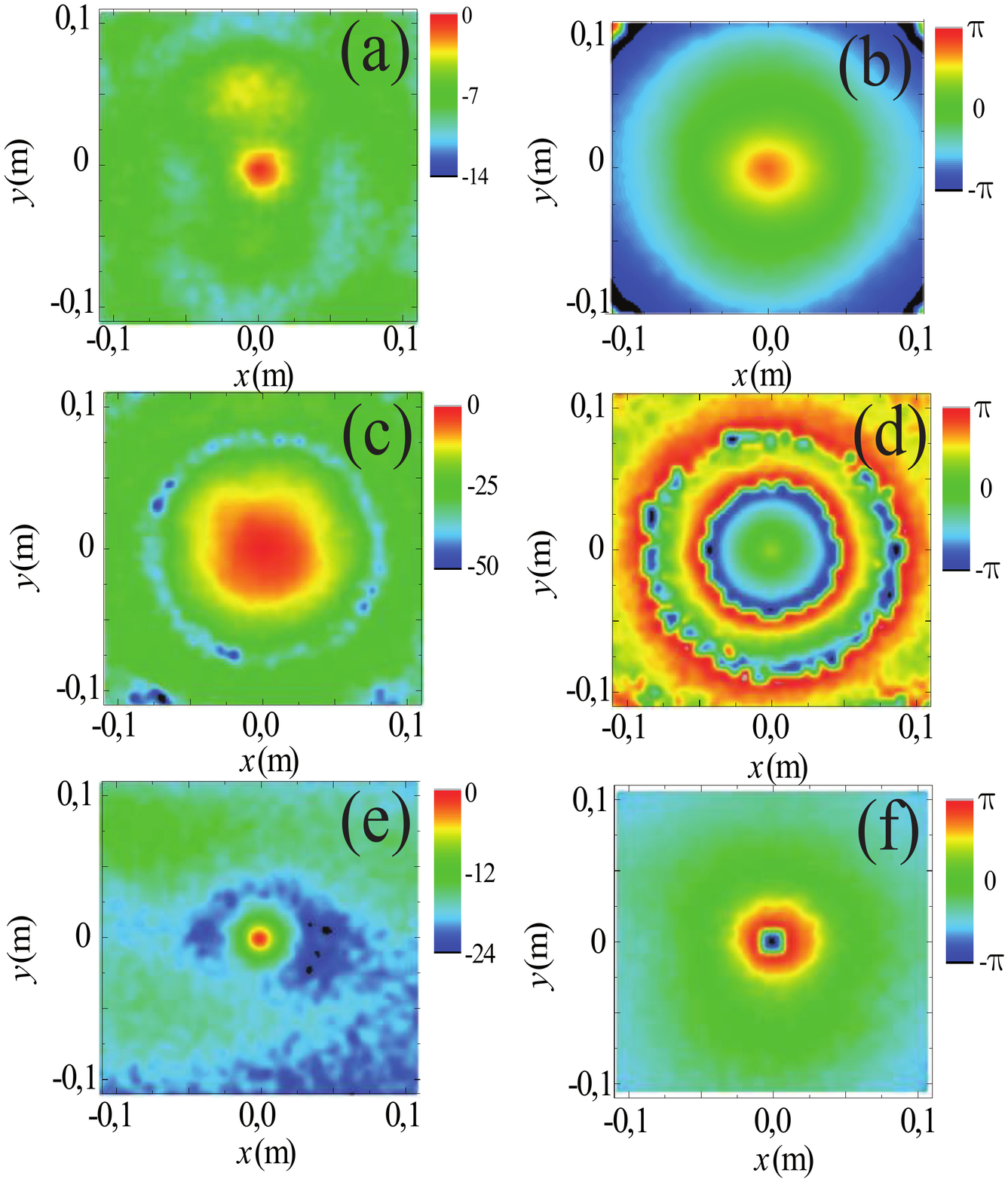}}
\caption{{Experimentally measured and numerically simulated (a,c) amplitude (in dB) and (b,d) phase (in radians)
 of the transmitted radiation, in the case when the source is shifted from the center of the device. 
 The frequency is  $f=820$ MHz. }} 
 \label{fig3}
\end{figure}

Figure~\ref{fig3} shows our experimental and numerical results for the transmitted radiation at the frequency $820$ MHz, for the case when the source 
is shifted along the horizontal axis towards the left edge of the structure. In the numerical modeling, we observe the formation of the standing mode 
formed by the surface waves, as predicted earlier~\cite{Finkfirst}. In our experiment, the amplitude and phase patterns are significantly blurred, and 
the presence of the standing wave is hardly distinguishable. This effect could be both due to the presence of losses in the dielectric host media and 
due to that all the wires are of slightly different lengths, which is crucial for the formation of standing wave. Moreover, sensitivity of the experimental 
measurements is  limited significantly by the dynamical range of the setup. In particular, at least $40$ dB have been lost due to the source and probe mismatch.

\begin{figure}[h]
\centerline{\includegraphics[width=0.95\columnwidth]{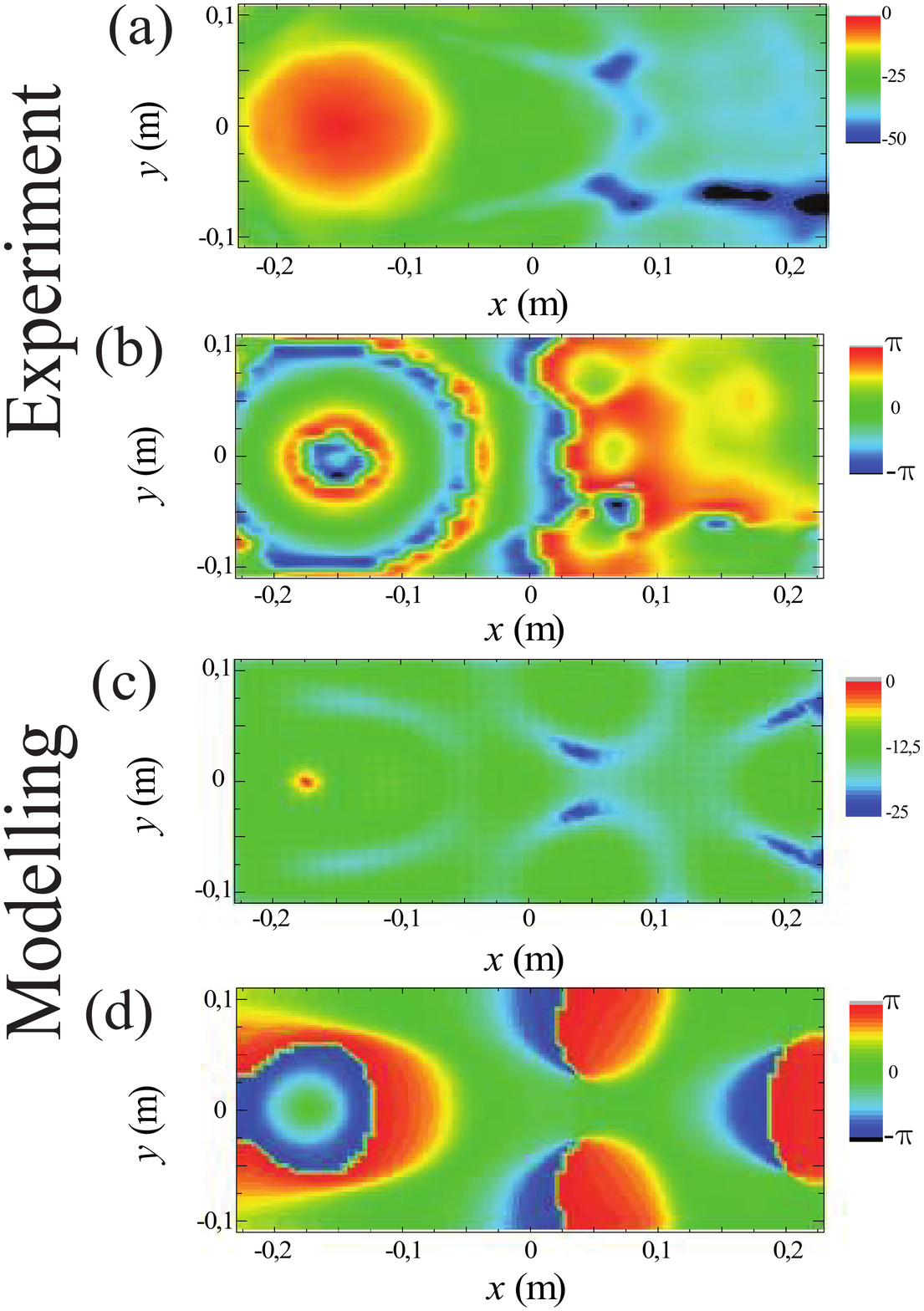}}
\caption{{Experimentally measured amplitude (left column, in dB) and phase (right column, in radians) of the transmitted radiation for different frequencies 
in the case when the source is placed in the center of the structure for the frequencies: (a,b) $f=760$ MHz, (c,d) $f=860$ MHz, and (e,f) $f=960$ MHz. }} \label{fig4}
\end{figure}

Figure~\ref{fig4} shows the amplitude and phase maps of the transmitted radiation for three different values of frequencies. We can observe a rapid phase 
change for the case of $860$ MHz. Furthermore, the amplitude map of the radiation at this frequency is characterized by a sharp dip at certain distance 
from the source. It is evident that for smaller and larger distances ($f=760$ MHz and $f=960$ Mhz), none of these distinctive features is observed.

\begin{figure}[h]
\centerline{\includegraphics[width=0.95\columnwidth]{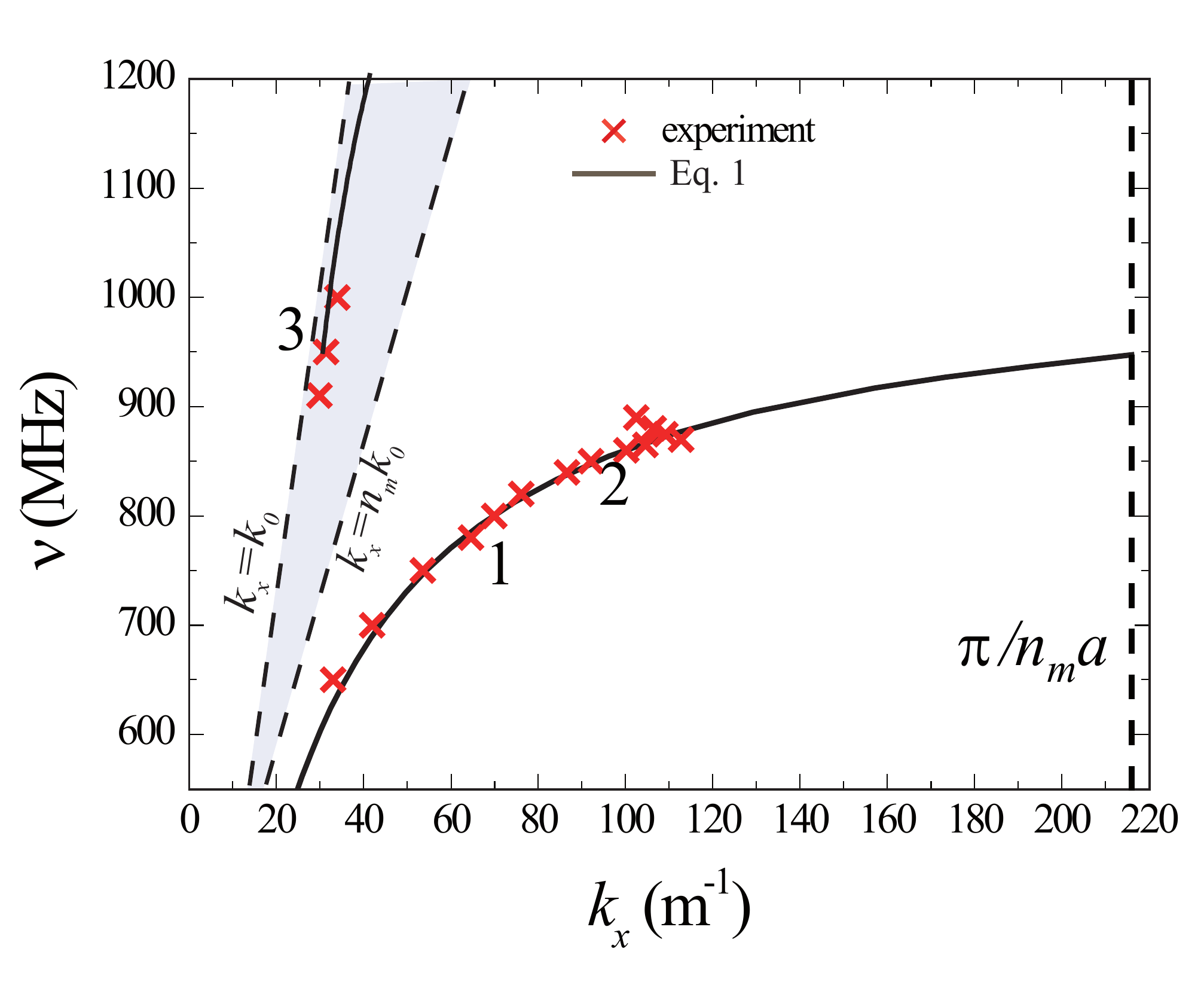}}
\caption{{Dispersion of guided modes extracted from the experimentally measured phase distribution (crosses). 
Solid curves show the analytical approximation for the dispersion relations. Blue area show the domain of 
existence of the guided modes in the corresponding conventional dielectric waveguide.}} 
\label{fig5}
\end{figure}

Next, we extract the waveguide numbers of the transmitted radiation from the phase maps. Assuming that the most part of the electromagnetic radiation 
is transferred in the guided modes, we assume the spatial dependence of the field in the transmitted wave to be $E\sim \exp[ik_{\rho} \rho]$, where $\rho$ 
is the distance from the source in the plane parallel to the structure surface, and $k_{\rho}$ is the wavenumber. Thus, the phase of the transmitted radiation 
should be linearly proportional to the wavenumber $k_{\rho}$. Therefore, it is possible to extract the dispersion of $k_{\rho}$ from the phase distribution 
map measured at different frequencies. Figure~\ref{fig5} shows the extracted values of the wavenumber together with the fitting with Eq.~\eqref{disprel}.  
We observe that the experimentally obtained dispersion relations agree well with the theoretical results. In the fitting, we have used the dielectric host 
media refractive index as a fitting parameter, since it was now known \textit{a priori}. The resulting value of $n_m=1.53$ is close to the typical values 
of the refractive index of the acetal for the considered frequency range.

We note that the result~\eqref{disprel} for the effective dispersion does not take into account the inherent periodicity of the metamaterial structure 
which limits the range of the allowed wavenumbers by the edge of the Brillouin zone $\pi n_m/a$, where $a$ is the lattice spacing. However, 
the measured wavenumbers do not reach this limiting value in experiment, and the dispersion curve bends in the middle of the band 
close to the value $\pi n_m/2a$. This bending of the dispersion curves is typical for surface plasmon polaritons in the presence of losses. 
The observed bending of the dispersion curves in our case originates from the losses in the dielectric host media and radiation losses in the structure.

In conclusion, we have studied the properties of waveguiding modes propagating in the slab of wire metamaterial taking into account 
dielectric permittivity of the host medium, and compared them with the corresponding properties of conventional waveguides.
We have studied experimentally propagating guided modes in a slab of wire metamaterial and measured their dispersion. 

This work was supported by the Ministry of Education and Science of Russian Federation (Grants No. 11.G34.31.0020, 14.B37.21.1649 and 14.B37.21.1941), the Dynasty Foundation, Russian Foundation for Basic Research (RFBR), and the Australian Research Council.

\end{document}